\documentclass[12pt]{article}

\usepackage[top=17mm, bottom=17mm, left=20mm, right=20mm]{geometry}
\usepackage{url, graphicx, color, varioref, amsfonts, longtable, float}
\newtheorem{theorem}{Theorem}[section]
\newtheorem{lemma}[theorem]{Lemma}

\newenvironment{proof}[1][Proof]{\begin{trivlist}
\item[\hskip \labelsep {\bfseries #1}]}{\end{trivlist}}

\newcommand{\qed}{\nobreak \ifvmode \relax \else
      \ifdim\lastskip<1.5em \hskip-\lastskip
      \hskip1.5em plus0em minus0.5em \fi \nobreak
      \vrule height0.75em width0.5em depth0.25em\fi}

\title{Fixed Phase Quantum Search Algorithm}

\author{Ahmed Younes\footnote {ayounes2@yahoo.com}\\
Department of Math. \& Comp. Science\\
Faculty of Science\\
Alexandria University\\
Alexandria, Egypt}

\begin{document}
\maketitle
\begin{abstract}
Building quantum devices using fixed operators is a must to simplify the hardware construction. Quantum search 
engine is not an exception. In this paper, a fixed phase quantum search algorithm that searches for $M$ matches in an unstructured search space of size $N$ will be presented. 
Selecting phase shifts of $1.91684\pi$ in the standard amplitude amplification will make the 
technique perform better so as to get probability of success at least 99.58$\%$ in 
$O\left( {\sqrt {{N \mathord{\left/ {\vphantom {N M}} \right.\kern-\nulldelimiterspace} M}} } \right)$ 
better than any know fixed operator quantum search algorithms. 
 The algorithm will be able to handle either a single match or multiple matches in the search space. 
 The algorithm will find a match in 
 $O\left( {\sqrt {{N \mathord{\left/ {\vphantom {N M}} \right.\kern-\nulldelimiterspace} M}} } \right)$ 
 whether the number of matches is known or not in advance.
\end{abstract}

\section{Introduction}       

In 1996, Lov Grover \cite{grover96} presented an algorithm that quantum mechanically searches 
an unstructured list assuming that a unique match exists in the list with quadratic speed-up 
over classical algorithms. To be able to define the target problem of this paper, 
we have to organize the efforts done by others in that field. 
The unstructured search problem targeted by Grover's original algorithm is deviated 
in the literature to the following four major problems:
\begin{itemize}
    \item Unstructured list with a unique match.
    \item Unstructured list with one or more matches, where the number of matches is known 
    \item Unstructured list with one or more matches, where the number of matches is unknown.
    \item Unstructured list with strictly multiple matches.
\end{itemize}

The efforts done in all the above cases, similar to Grover's original work, used quantum parallelism by 
preparing superposition that represents all the items in the list. The superposition could be uniform or arbitrary.
 The techniques used in most of the cases to amplify the amplitude(s) of the required state(s) 
 have been generalized to an amplitude amplification technique that iterates the operation $UR_s \left( \phi  \right)U^\dag  R_t \left( \varphi  \right)$, on $U\left| s \right\rangle$ where 
$U$ is unitary operator, $R_s \left( \phi  \right) = I - (1 - e^{i\phi } )\left| s \right\rangle \left\langle s \right|$,
$R_t \left( \varphi  \right) = I - (1 - e^{i\varphi } )\left| t \right\rangle \left\langle t \right|$, 
$\left| s \right\rangle$ is the initial state of the system, $\left| t \right\rangle$ represents 
the target state(s) and $I$ is the identity operator.

Grover's original algorithm replaces $U$ be $W$, where $W$ is the Walsh-Hadamard transform, 
prepares the superposition $W\left| 0 \right\rangle$ (uniform superposition) and iterates 
$WR_s \left( \pi  \right)WR_t \left( \pi  \right)$ for $O\left( {\sqrt N } \right)$, 
where $N$ is the size of the list, which was shown be optimal to get the highest 
probability with the minimum number of iterations \cite{Zalka99}, 
such that there is only one match in the search space.
 
In \cite{Grover98a,Jozsa99,Gal00,long01,BK02}, Grover's algorithm is generalized by showing that 
$U$ can be replaced by almost any arbitrary superposition and the phase shifts $\phi$ and $\varphi$ 
can be generalized to deal with the arbitrary superposition and/or to increase the probability of 
success even with a factor increase in the number of iterations to still run in $O(\sqrt{N})$.
These give a larger class of algorithms for amplitude amplification using variable operators 
from which Grover's algorithm was shown to be a special case.

In another direction, work has been done trying to generalize Grover's algorithm with a uniform 
superposition for known number of multiple matches in the search space 
\cite{boyer96,Chen99,Chen00a,Chen00b}, 
where it was shown that the required number of iterations is approximately 
${\pi}/{4}\sqrt {{N}/{M}}$ for small ${M}/{N}$, where $M$ is the number of matches. 
The required number of iterations will increase for $M>{N}/{2}$, i.e. the problem will be harder 
where it might be excepted to be easier \cite{nc00a}. Another work has been done 
for known number of multiple matches with arbitrary superposition and 
phase shifts \cite{Mosca98,Biron98,Brassard00,hoyer00,Li01} where the same problem 
for multiple matches occurs. In \cite{Brassard98,Mosca98,Brassard00}, 
a hybrid algorithm was presented to deal with this problem 
by applying Grover's fixed operators algorithm for ${\pi}/{4}\sqrt {{N}/{M}}$ 
times then apply one more step using specific $\phi$ and $\varphi$ according to the knowledge of 
the number of matches $M$ to get the solution with probability close to certainty. 
Using this algorithm will increase the hardware cost since we have to build one more 
$R_s$ and $R_t$ for each particular $M$. For the sake of practicality, the operators should be fixed for any 
given $M$ and are able to handle the problem with high probability whether or not $M$ is known in advance. 
In \cite{Younes03d,Younes04a}, Younes et al presented an algorithm that exploits entanglement and partial diffusion 
operator to perform the search and can perform in case of either a single match or 
multiple matches where the number of matches is known or not \cite{Younes04a} 
covering the whole possible range, i.e. $1 \le M \le N$. 
Grover described this algorithm as the best quantum search algorithm \cite{Groverbest}. 
It can be shown that we can get the same probability of success of \cite{Younes03d} using amplitude 
amplification with phase shifts $\phi=\varphi=\pi/2$, 
although the amplitude amplification mechanism will be different. 
The mechanism used to manipulate the amplitudes could be useful in many applications, 
for example, superposition preparation and error-correction. 

For unknown number of matches, an algorithm for estimating the number of matches 
({\it quantum counting algorithm}) was presented \cite{Brassard98,Mosca98}. 
In \cite{boyer96}, another algorithm was presented to find a match even if the number of matches is unknown 
which will be able to work if $M$ lies within the range $1\le M \le 3N/4$ \cite{Younes04a}.
 
For strictly multiple matches, Younes et al \cite{Younes03c} presented an algorithm which works 
very efficiently only in case of {\it multiple matches} within the search space that splits the solution states over more states, inverts the sign of half of them (phase shift of -1) and keeps the other half unchanged every iteration. This will keep the mean of the amplitudes to a minimum for multiple matches. The same result was rediscovered by Grover using amplitude amplification with phase shifts $\phi=\varphi=\pi/3$ \cite{grover150501}, in both algorithms the behavior will be similar to the classical algorithms in the worst case.

In this paper, we will propose a fixed phase quantum search algorithm that runs in 
$O\left( {\sqrt {{N \mathord{\left/ {\vphantom {N M}} \right. \kern-\nulldelimiterspace} M}}}\right)$. 
This algorithm is able to handle the range $1\le M \le N$ for both known and unknown number of matches 
more reliably than known fixed operator quantum search algorithms that target this case.

The plan of the paper is as follows: Section 2 introduces the general definition of the target unstructured search problem. Section 3 presents the algorithm for both known and unknown number of matches. The paper will end up with a general conclusion in Section 4.

\section{Unstructured Search Problem}

Consider an unstructured list $L$ of $N$ items. For simplicity and without loss of generality we will assume that $N = 2^n$ for some positive integer $n$. Suppose the items in the list are labeled with the integers $\{0,1,...,N - 1\}$, and consider a function (oracle) $f$ which maps an item $i \in L$ to either 0 or 1 according to some properties this item should satisfy, i.e. $f:L \to \{ 0,1\}$. The problem is to find any $i \in L$ such that $f(i) = 1$ assuming that such $i$ exists in the list. In conventional computers, solving this problem needs $O\left({N}/{M}\right)$ calls to the oracle (query),where $M$ is the number of items that satisfy the oracle.

\section{Fixed Phase Algorithm}
\subsection{Known Number of Matches}

Assume that the system is initially in state $\left| s \right\rangle  = \left| 0 \right\rangle$. Assume that $\sum\nolimits_i {^{'}} $ denotes a sum over $i$ which are desired matches, and $\sum\nolimits_i {^{''}} $ denotes a sum over $i$ which are undesired items in the list. So, Applying $U\left| s \right\rangle$ we get,

\begin{equation}
\left| \psi^{(0)}  \right\rangle=U\left| s \right\rangle  = \frac{1}{{\sqrt N }}\sum\limits_{i = 0}^{N - 1} {^{'} \left| i \right\rangle }  
+ \frac{1}{{\sqrt N }}\sum\limits_{i = 0}^{N - 1} {^{''} \left| i \right\rangle }, 
\end{equation}
\noindent
where $U=W$ and the superscript in $\left| \psi^{(0)}  \right\rangle$ represents the iteration number.

Let $M$ be the number of matches, 
$\sin (\theta ) = \sqrt {{M \mathord{\left/ {\vphantom {M N}} \right.\kern-\nulldelimiterspace} N}}$ 
and $0 < \theta  \le \pi /2$, then the system can be re-written as follows,

\begin{equation}
\left| \psi^{(0)}  \right\rangle  = \sin (\theta )\left| {\psi _1 } \right\rangle  + \cos (\theta )\left| {\psi _0 } \right\rangle,
\end{equation}

\noindent
where $\left| {\psi _1 } \right\rangle=\left| {t } \right\rangle$ represents the matches subspace and $\left| {\psi _0 } \right\rangle$ represents the non-matches subspace. 

Let $D=UR_s \left( \phi  \right)U^\dag  R_t \left( \varphi  \right)$, $R_s \left( \phi  \right) = I - (1 - e^{i\phi } )\left| s \right\rangle \left\langle s \right|$, $R_t \left( \varphi  \right) = I - (1 - e^{i\varphi } )\left| t \right\rangle \left\langle t \right|$ and set $\phi=\varphi$ as the best choice \cite{hoyer00}. 
Applying $D$ on $\left| \psi^{(0)}  \right\rangle$ we get,

%From now on, we will assume that $\phi=\varphi=1.91684\pi$. 

\begin{equation}
\left| {\psi ^{(1)} } \right\rangle  = D\left| {\psi ^{(0)} } \right\rangle  = a_1 \left| {\psi _1 } \right\rangle  + b_1 \left| {\psi _0 } \right\rangle ,
\end{equation}

\noindent
such that,
\begin{equation}
a_1  = \sin (\theta )(2\cos \left( \delta  \right)e^{i\phi }  + 1),
\end{equation}
\begin{equation}
b_1  = e^{i\phi } \cos (\theta )(2\cos \left( \delta  \right) + 1),
\end{equation} 
\noindent
where $\cos \left( \delta  \right) = 2\sin ^2 (\theta )\sin ^2 ({\textstyle{\phi  \over 2}}) - 1$.

Let $q$ represents the required number of iterations to get a match with the 
highest possible probability. After $q$ applications of $D$ on $\left| {\psi ^{(0)} } \right\rangle$ we get,

\begin{equation}
\left| {\psi ^{(q)} } \right\rangle  = D^{q}\left| {\psi ^{(0)} } \right\rangle  = a_q \left| {\psi _1 } \right\rangle  + b_q \left| {\psi _0 } \right\rangle ,
\end{equation}

\noindent
such that, 
\begin{equation}
\label{aqeqn}
a_q  = \sin (\theta )\left( {e^{iq\phi } U_q \left( y \right) + e^{i(q - 1)\phi } U_{q - 1} \left( y \right)} \right), 
\end{equation}

\begin{equation}
b_q =  \cos (\theta )e^{i(q - 1)\phi } \left( {U_q \left( y \right) + U_{q - 1} \left( y \right)} \right),
\end{equation} 

\noindent  
where $y=cos(\delta)$ and $U_q$ is the Chebyshev polynomial of the second kind defined as follows, 

\begin{equation}
 U_q \left( y \right) = \frac{{\sin \left( {\left( {q + 1} \right)\delta } \right)}}{{\sin \left( \delta  \right)}}.
\end{equation}

Let $P_s^q$ represents the probability of success to get a match after $q$ 
iterations and $P_{ns}^q$ is the probability not to get a match after applying measurement, 
so $P_s^q  = \left| {a_q } \right|^2$ and $P_{ns}^q  = \left| {b_q } \right|^2$ 
such that $P_s^q  + P_{ns}^q  = 1$. To calculate the required number of iterations $q$ to 
get a match with certainty, one the following two approaches might be followed:

\begin{itemize}
\item {Analytically}. The usual approach used in the literature when the number of matches $M$  
is known in advance is to equate $P_s^q$ to 1 or $P_{ns}^q$ to 0 and then find an algebraic 
formula that represents the required number of iterations, as well as, the phase shifts $\phi$ and 
$\varphi$ in terms on $M$. Using this approach is not possible 
for the case that the phase shifts should be fixed for an arbitrary $M$ such that $1 \le M \le N$ as shown 
in the following theorem.

\begin{theorem}[No Certainty Principle]

Let $D$ be an amplitude amplification operator such that 
$D=UR_s \left( \phi  \right)U^\dag  R_t \left( \varphi  \right)$, 
where $U$ is unitary operator, 
$R_s \left( \phi  \right) = I - (1 - e^{i\phi } )\left| s \right\rangle \left\langle s \right|$,
$R_t \left( \varphi  \right) = I - (1 - e^{i\varphi } )\left| t \right\rangle \left\langle t \right|$, 
$\left| s \right\rangle$ is the initial state of the system, $\left| t \right\rangle$ represents 
the target state(s) and $I$ is the identity operator. Let $D$ performs on a system initially set to 
$U\left| s \right\rangle$. If the phase shifts $\phi$ and $\varphi$ 
should be fixed, then iterating $D$ an arbitrary number of times will not find a match with certainty 
for an arbitrary known number of matches $M$  such that $1 \le M \le N$.

\begin{proof}

To prove this theorem, we will use the usual approach, i.e. start with $P_s^q=1$ or $P_{ns}^q=0$ and 
calculate the required number of iterations $q$. 

Since $P_s^q  = \left| {a_q } \right|^2$ and from Eqn.\ref{aqeqn}, we can re-write 
$P_s^q$ as follows setting $\phi=\varphi$ as the best choice \cite{hoyer00},

\begin{equation}
\label{psth}
P_s^q  = \frac{{\sin ^2 \left( \theta  \right)}}{{\sin ^2 \left( \delta  \right)}}\left( {1 - \cos \left( \delta  \right)\cos \left( {\left( {2q + 1} \right)\delta } \right) + 2\cos \left( \phi  \right)\sin \left( {\left( {q + 1} \right)\delta } \right)\sin \left( {q\delta } \right)} \right).
\end{equation}

Setting $P_s^q=1$ and using simple trigonometric identities we get, 
$q = \frac{{ - 1}}{2}$, i.e. the required number of iterations is independent 
of $M$, $\phi$ and $\varphi$, and represents an impossible value for a required 
number of iterations.

\end{proof}
\end{theorem}

\item {Direct Search.} The alternative approach used in this paper is to empirically 
assume an algebraic form for the required number of iterations that satisfy 
the quadratic speed-up of the known quantum search algorithms and use 
a computer program to search for the best phase shift $\phi$ that satisfy the condition,

\begin{equation}
\label{maxmin}
\max \left( {\min \left( {P_s^q (\phi )} \right)} \right) \,\, such\,\,that\,\, 0 \le \phi  \le 2\pi\,\, and,\,\, 1 \le M \le N.
\end{equation}

\noindent
i.e. find the value of $\phi$ that maximize the minimum value of $P_s^q$ over the 
range $1 \le M \le N$. 
  
\end{itemize}

\begin{figure}[t]
\centerline{\includegraphics[width=5.0in,height=3.75in]{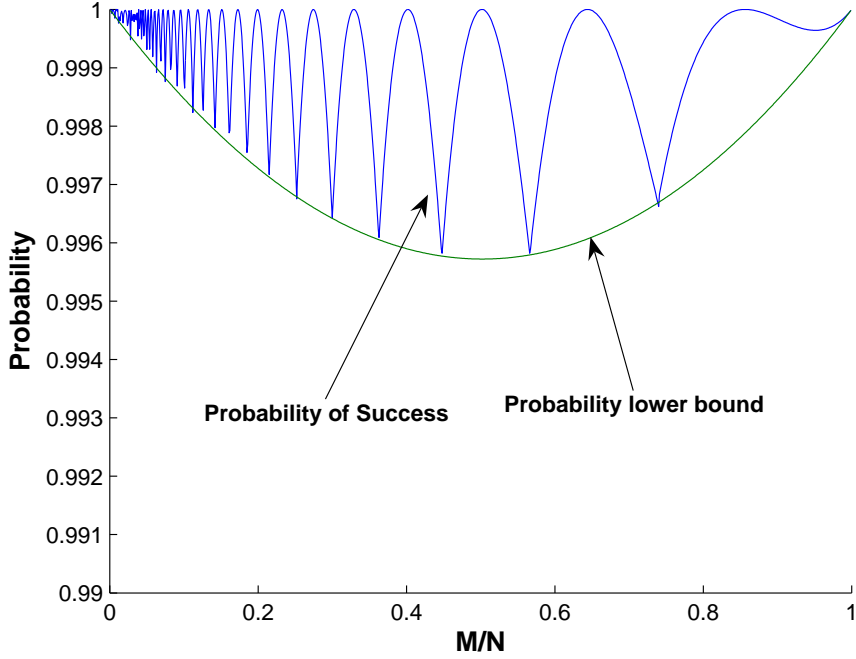}}
\caption{The probability of success the proposed algorithm after the required number of iterations.}
\label{problbeps}
\end{figure}

\begin{figure}[t]
\centerline{\includegraphics[width=5.0in,height=3.75in]{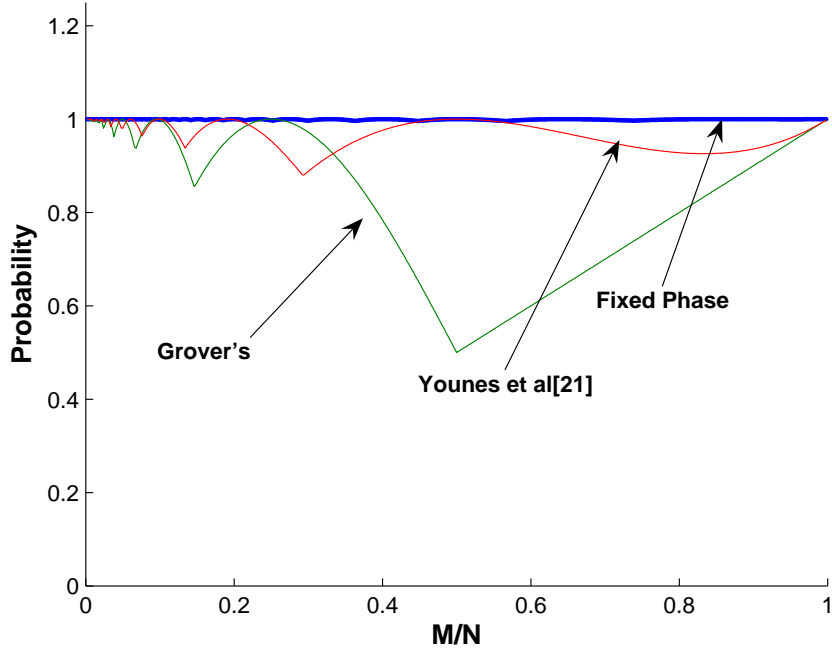}}
\caption{The probability of success of Grover's algorithm, Younes et al algorithm \cite{Younes03d} 
and the proposed algorithm after the required number of iterations.}
\label{compPs}
\end{figure}

Assume that $q = \left\lfloor {{\textstyle{\phi  \over {\sin (\theta)}}}} \right\rfloor  = O\left( {\sqrt {{\textstyle{N \over M}}} } \right)$.  
Using this form for $q$, a computer program has been written using C language 
to find the best $\phi$ with precision $10^{ - 15}$ 
that satisfy the conditions shown in Eqn.~\ref{maxmin}. The program shows that using $\phi=6.021930660106538\approx1.91684\pi$, the minimum probability of 
success will be at least $99.58\%$ compared with 87.88 $\%$ 
for Younes et al \cite{Younes04a} and 50$\%$ for the original Grover's algorithm \cite{boyer96} 
as shown in Fig.~\ref{compPs}. To prove these results, using $\phi=1.91684\pi$, the lower bound 
for the probability of success is as follows as shown in Fig.~\ref{problbeps}.

\begin{equation}
\begin{array}{l}
 P_s^q  = \frac{{\sin ^2 \left( \theta  \right)}}{{\sin ^2 \left( \delta  \right)}}\left( {1 - \cos \left( \delta  \right)\cos \left( {\left( {2q + 1} \right)\delta } \right) + 2\cos \left( \phi  \right)\sin \left( {\left( {q + 1} \right)\delta } \right)\sin \left( {q\delta } \right)} \right) \\ 
  \,\,\,\,= \frac{{\sin ^2 \left( \theta  \right)}}{{\sin ^2 \left( \delta  \right)}}\left( {1 - \cos \left( \delta  \right)\cos \left( {\left( {2q + 1} \right)\delta } \right) + \cos \left( \phi  \right)\cos \left( \delta  \right) - \cos \left( \phi  \right)\cos \left( {\left( {2q + 1} \right)\delta } \right)} \right) \\ 
  \,\,\,\,\ge \frac{{\sin ^2 \left( \theta  \right)}}{{\sin ^2 \left( \delta  \right)}}\left( {1 + \cos ^2 \left( \delta  \right)+ 2 \cos \left( \phi  \right)\cos \left( \delta  \right)}\right) \ge 0.9958. \\ 
 \end{array}
\end{equation}

\noindent
where, $\cos \left( \delta  \right) = 2\sin ^2 (\theta )\sin ^2 ({\textstyle{\phi  \over 2}}) - 1$,
 $0<\theta\le \pi/2$, and ${\cos \left( {\left( {2q + 1} \right)\delta } \right)}\le-cos(\delta)$.
 
\subsection{Unknown Number of Matches}
 
In case we do not know the number of matches $M$ in advance, 
we can apply the algorithm shown in \cite{boyer96} for $1 \le M\le N$ 
by replacing Grover's step with the proposed algorithm. 
The algorithm can be summarized as follows,

\begin{itemize}
\item[1-] Initialize $m=1$ and $\lambda= {8}/{7}$. (where $\lambda$ can take any value between 
1 and ${4}/{3}$)
\item[2-]Pick an integer $j$ between 0 and $m-1$ in a uniform random manner.
\item[3-]Run $j$ iterations of the proposed algorithm on the state $\left| {\psi ^{\left( 0 \right)} } \right\rangle$:
 
\begin{equation}
\left| {\psi ^{\left( j \right)} } \right\rangle  = D^j \left| {\psi ^{\left( 0 \right)} } \right\rangle.
\end{equation}

\item[4-]Measure the register $\left| {\psi ^{\left( j \right)} } \right\rangle $ and assume $i$ is the output.
\item[5-]If $f(i)=1$, then we found a solution and exit.
\item[6-]Set $m=min\left( \lambda m,\sqrt{N}\right)$ and go to step 2.
\end{itemize} 

\noindent
where $m$ represents the range of random numbers (step 2), 
$j$ represents the random number of iterations (step3), 
and $\lambda$ is a factor used to increase the range of random numbers 
after each trial (step 6). 

For the sake of simplicity and to be able to compare the performance of this algorithm with that shown in \cite{boyer96}, we will try to follow the same style of analysis used in \cite{boyer96}. Before we construct the analysis, we need the following lemmas.

\begin{lemma}
\label{lemch5}
For any positive integer $m$ and real numbers $\theta$, $\delta$ such that 
$\cos \left( \delta  \right) = c\sin ^2 (\theta ) - 1$,  
$0 < \theta  \le \pi /2$ where $c=2\sin ^2 ({\textstyle{\phi  \over 2}})$ is a constant, 
\[
\sum\limits_{q = 0}^{m - 1} {\sin ^2 \left( {\left( {q + 1} \right)\delta } \right) + \sin ^2 \left( {q\delta } \right) = m - \frac{{\cos \left( \delta  \right)\sin \left( {2m\delta } \right)}}{{2\sin \left( \delta  \right)}}} .
\]

\begin{proof} 
By mathematical induction. %$\,\,\,\,\Box$
\end{proof}
\end{lemma}

\begin{lemma}
\label{lemch6}
For any positive integer $m$ and real numbers $\theta$, $\delta$ such that 
$\cos \left( \delta  \right) = c\sin ^2 (\theta ) - 1$,  
$0 < \theta  \le \pi /2$ where $c=2\sin ^2 ({\textstyle{\phi  \over 2}})$ is a constant, 
\[
\sum\limits_{q = 0}^{m - 1} {\sin \left( {\left( {q + 1} \right)\delta } \right)} \sin \left( {q\delta } \right) = \frac{m}{2}\cos \left( \delta  \right) - \frac{{\sin \left( {2m\delta } \right)}}{{4\sin \left( \delta  \right)}}.
\]

\begin{proof} 
By mathematical induction. %$\,\,\,\,\Box$
\end{proof}
\end{lemma}

\begin{lemma}

Assume $M$ is the unknown number of matches such that $1\le M \le N$. Let $\theta$, $\delta$ be real numbers such that $\cos \left( \delta  \right) = 2\sin ^2 (\theta )\sin ^2 ({\textstyle{\phi  \over 2}}) - 1$, $\sin ^2 (\theta ) = M/N$, $\phi=1.91684\pi$ and $0 < \theta  \le \pi /2$. Let $m$ be any positive integer. Let $q$ be any integer picked in a uniform random manner between 0 and $m-1$. Measuring the register after applying $q$ iterations of the proposed algorithm starting from the initial state, the probability $P_m$ of finding a solution is as follows,

\[
P_m  = \frac{1}{c\left({1 - \cos \left( \delta  \right)}\right)}\left( {1 + \cos \left( \delta  \right)\cos \left( \phi  \right) - \frac{{\left( {\cos \left( \delta  \right) + \cos \left( \phi  \right)} \right)\sin \left( {2m\delta } \right)}}{{2m\sin \left( \delta  \right)}}} \right),
\]

\noindent
where $c = 2\sin ^2 ({\textstyle{\phi  \over 2}})$, then $P_m\ge1/4$ for $m \ge 1/\sin \left( \delta  \right)$ and small $M/N$.

\begin{proof} 

The average probability of success when applying $q$ iterations of the proposed algorithm when  
$0\le q \le m$ is picked in a uniform random manner is as follows,

\[
\begin{array}{l}
 P_m  = \frac{1}{m}\sum\limits_{q = 0}^{m - 1} {P_s^q }  \\ 
 \,\,\,\,\,\,\, = \frac{{\sin ^2 \left( \theta  \right)}}{{m\sin ^2 \left( \delta  \right)}}\sum\limits_{q = 0}^{m - 1} {\left( {\sin ^2 \left( {\left( {q + 1} \right)\delta } \right) + \sin ^2 \left( {q\delta } \right) + 2\cos \left( \phi  \right)\sin \left( {\left( {q + 1} \right)\delta } \right)\sin \left( {q\delta } \right)} \right)}  \\ 
 \,\,\,\,\,\,\, = \frac{{\sin ^2 \left( \theta  \right)}}{{m\sin ^2 \left( \delta  \right)}}\left( {m - \frac{{\cos \left( \delta  \right)\sin \left( {2m\delta } \right)}}{{2\sin \left( \delta  \right)}} + \cos \left( \phi  \right)\cos \left( \delta  \right) - \frac{{\cos \left( \phi  \right)\sin \left( {2m\delta } \right)}}{{2\sin \left( \delta  \right)}}} \right) \\
 \,\,\,\,\,\,\, = \frac{1}{c\left({1 - \cos \left( \delta  \right)}\right)}\left( {1 + \cos \left( \delta  \right)\cos \left( \phi  \right) - \frac{{\left( {\cos \left( \delta  \right) + \cos \left( \phi  \right)} \right)\sin \left( {2m\delta } \right)}}{{2m\sin \left( \delta  \right)}}} \right),\\
 \end{array}
\]

\noindent
If $m \ge 1/\sin \left( \delta  \right)$ and $M \ll N$ then $ \cos\left(\delta\right)\approx -1$, so,

\[
P_m  \ge \frac{1}{{2c}}\left( {1 - \cos \left( \phi  \right) - \frac{{\left( {\cos \left( \phi  \right) - 1} \right)\sin \left( {2m\delta } \right)}}{2}} \right) \ge \frac{1}{{2c}}\left( {1 - \cos \left( \phi  \right) - \frac{{\left( {1 - \cos \left( \phi  \right)} \right)}}{2}} \right) = 0.25 \\ 
\]

\noindent
where $-1\le\sin \left( {2m\delta } \right) \le 1$ for $0 < \theta  \le \pi /2$. %$\,\,\,\,\Box$

\end{proof}
\end{lemma}

We calculate the total expected number of iterations as done in Theorem 3 in \cite{boyer96}. Assume that $m_q  \ge 1/\sin \left( \delta  \right)$, and $v_q  = \left\lceil {\log _\lambda  m_q } \right\rceil $. 
Notice that, $m_q  = O\left( {\sqrt {N/M} } \right)$ for $1 \le M \le N$, then:  

\begin{itemize}

\item[1-] The total expected number of iterations to reach the critical stage, i.e. when $m\ge m_q $:

\begin{equation}
\frac{1}{2}\sum\limits_{v = 1}^{v_q } {\lambda ^{v - 1} }  \le \frac{1}{{2\left( {\lambda  - 1} \right)}}m_q  = 3.5m_q.
\end{equation}
 
\item[2-] The total expected number of iterations after reaching the critical stage:

\begin{equation}
\frac{1}{2}\sum\limits_{u = 0}^\infty  {\left( {\frac{3}{4}} \right)^u \lambda ^{v_q  + u}  = \frac{1}{{2\left( {1 - 0.75\lambda } \right)}}} m_q  = 3.5m_q.
\end{equation}

\end{itemize}

The total expected number of iterations whether we reach to the critical stage or not is $7m_q$ which is in $O(\sqrt{N/M})$ for $1\le M\le N$.

\begin{figure}[t]
%\centerline{\includegraphics[width=4.50in,height=3.25in]{Figs/764mqg.eps}}
\centerline{\includegraphics[width=5.0in,height=3.75in]{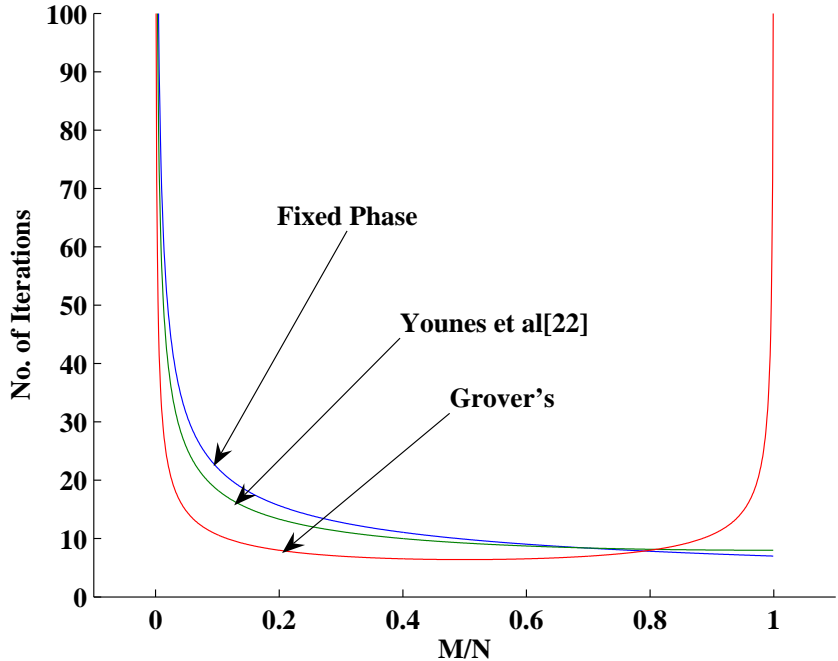}}
\caption{The actual behavior of the functions representing the total expected number of iterations 
for Grover's algorithm, Younes et al algorithm \cite{Younes04a} and the proposed algorithm 
taking $\lambda=8/7$, where the number of iterations is the flooring of the values (step function).} 
\label{78mqg}
\end{figure}

When this algorithm employed Grover's algorithm, and based on the condition 
$m_G  \ge 1/\sin \left( {2\theta _G } \right)= O\left( {\sqrt {N/M} } \right)$ for $M\le {3N}/{4}$,the total expected number of iterations is approximately $8m_G$ for $1 \le M\le {3N}/{4}$. Employing the proposed algorithm instead, and based on the condition $m_q  \ge 1/\sin \left( {\delta} \right)= O\left( {\sqrt {N/M} } \right)$,the total expected number of iterations is approximately $7m_q$ for $1\le M\le N$, i.e. the algorithm will be able to handle the whole range, since $m_q$ will be able to act as a lower bound for $q$ over $1\le M\le N$. Fig.~\ref{78mqg} compares between the total expected number of iterations for Grover's algorithm, Younes et al algorithm \cite{Younes04a} and the Fixed Phase algorithm taking $\lambda  = 8/7$.

%\begin{figure}[t]
%\centerline{\includegraphics[width=4.00in,height=3.0in]{Figs/7m8m.eps}}
%\caption{Behavior of the total expected number of iterations for Grover's algorithm $8 m_G$ and the 
%proposed algorithm $7 m_q$.} 
%\label{PDqmq}
%\end{figure}

\section{Conclusion}
To be able to build a practical search engine, the engine should be constructed from fixed operators that can handle the whole possible range of the search problem, i.e. whether a single match or multiple matches exist in the search space. It should also be able to handle the case where the number of matches is unknown. The engine should perform with the highest possible probability after performing the required number of iterations.

In this paper, a fixed phase quantum search algorithm is presented. It was shown that selecting the phase shifts to $1.91684\pi$ could enhance the searching process so as to get a solution with probability at least 99.58$\%$. The algorithm still achieves the quadratic speed up of Grover's original algorithm. It was shown that Younes et al algorithm \cite{Younes04a} might perform better in case the number of matches is unknown, although the presented algorithm might scale similar with an acceptable delay. i.e. both run in $O\left( {\sqrt {{N \mathord{\left/ {\vphantom {N M}} \right. \kern-\nulldelimiterspace} M}} } \right)$. In that sense, the Fixed Phase algorithm can act efficiently in all the possible classes of the unstructured search problem.

\bibliography{pi1825}
\bibliographystyle{plain}

\end{document}